| | |
|---|---|
| Title | Plasma-activation of tap water using DBD for agronomy applications: Identification and quantification of long lifetime chemical species and production/consumption mechanisms |
| Authors | F. Judée[a], S. Simon[a], C. Bailly[b], T. Dufour[a] |
| Affiliations | [a] LPP, UPMC Univ. Paris 06, Sorbonne Universités, CNRS, Ecole Polytech., Univ. Paris-Sud, Observatoire de Paris, Université Paris-Saclay, PSL Research University, 4 Place Jussieu, 75252 Paris, France<br>[b] Sorbonne Universités, UPMC Univ Paris 06, CNRS, Institut de Biologie Paris-Seine (IBPS), UMR 7622, Biologie du développement, F-75005, Paris, France |
| Ref. | Water Research 133 (2018) 47-59 |
| DOI | https://doi.org/10.1016/j.watres.2017.12.035 |
| Abstract | Cold atmospheric plasmas are weakly ionized gases that can be generated in ambient air. They produce energetic species (e.g. electrons, metastables) as well as reactive oxygen species, reactive nitrogen species, UV radiations and local electric field. Their interaction with a liquid such as tap water can hence change its chemical composition. The resulting "plasma-activated liquid" can meet many applications, including medicine and agriculture.<br>Consequently, a complete experimental set of analytical techniques dedicated to the characterization of long lifetime chemical species has been implemented to characterize tap water treated using cold atmospheric plasma process and intended to agronomy applications. For that purpose, colorimetry and acid titrations are performed, considering acid-base equilibria, pH and temperature variations induced during plasma activation. 16 species are quantified and monitored: hydroxide and hydronium ions, ammonia and ammonium ions, orthophosphates, carbonate ions, nitrite and nitrate ions and hydrogen peroxide. The related consumption/production mechanisms are discussed. In parallel, a chemical model of electrical conductivity based on Kohlrausch's law has been developed to simulate the electrical conductivity of the plasma-activated tap water (PATW). Comparing its predictions with experimental measurements leads to a narrow fitting, hence supporting the self-sufficiency of the experimental set, i.e. the fact that all long lifetime radicals of interest present in PATW are characterized.<br>Finally, to evaluate the potential of cold atmospheric plasmas for agriculture applications, tap water has been daily plasma-treated to irrigate lentils seeds. Then, seedlings lengths have been measured and compared with untreated tap water, showing an increase as high as 34.0% and 128.4% after 3 days and 6 days of activation respectively. The interaction mechanisms between plasma and tap water are discussed as well as their positive synergy on agronomic results. |

# 1. Introduction

Civilization is founded on agriculture, remaining as important today as its beginning 10,000 years ago. Even if mechanization, technological innovations and chemicals have ensured higher productivity in regard of the two last centuries, modern agriculture must face new challenges today. The United Nations Food and Agriculture Organization (FAO) has indicated that global food shortages will become three times more likely owing to climate change and the rapid development of urbanization, industrialization and world population. In parallel, micropollutants (food additives, industrial chemicals, pesticides, pharmaceuticals and personal care products) constitute a class of hazardous products which rise major concerns. Despite their very low concentrations (detectable in the ng/L-mg/L range), many studies have shown their harmful effects on the environment (Milla et al., 2011; Rizzo et al., 2013) and in particular on agriculture (Jampeetong and Brix, 2009; Roosta et al., 2009).

According to the FAO, one of the most viable processes to limit food shortages is to increase crop yields, so far mainly limited by seed surface and by water and soil contamination with bacteria, microorganisms, fungi and various chemical compounds. There are also environmental concerns like insect pests, adverse weather conditions or human as storage and fertilization. Among the solutions considered to face such challenges, cold atmospheric plasmas (CAP) appear more and more as an innovative and ecofriendly approach. CAP are weakly ionized gases that can be generated in ambient air using simple devices called dielectric barrier discharges (DBD) reactors and that can easily be coupled with green electricity. Applying a high voltage between the two electrodes e with at least one of them covered by a dielectric barrier e enables the production of an intense electric field which in turn ionizes the gas (e.g. ambient air) confined in the interelectrode gap as described in Fig. 1 a. The dielectric barrier is of major importance as it prevents any arc formation and permits only the generation of cold microdischarges. As a consequence, energetic species (e.g. electrons, metastables) are generated as well as reactive oxygen species including atomic oxygen, hydroxyl radicals, ozone, but also radical ions, namely $O_2^+$, $O_2^-$, $O_3^-$ (Marotta et al., 2012).

DBD reactors can be utilized for the treatment (also called activation) of liquids: the gaseous radicals can directly interplay at the liquid interface or even diffuse in depth, leading to complex chemical mechanisms that are still being deciphered. In addition, the CAP can generate transient electric field, flow kinetics and heating effects which can directly impact on the chemical mechanisms. Such plasma-liquid interactions have recently drawn considerable attention in the field of water treatment: numerous studies have indeed shown the effectiveness of plasma DBD activation to decontaminate water, e.g. removing pesticides (Vanraes et al., 2017), phenols (Marotta et al., 2012) or pharmaceutical compounds (Magureanu et al., 2015). Other methods than the direct interaction between plasma and water are also used. Examples include the use of water drop plasma for Ibuprofen removal (Wang et al., 2017) or plasmas directly generated in water to produce "plasma bubbles" for E. coli inactivation (Ma et al., 2017b) and degradation process of chitosan (Ma et al., 2017a).

At the same time, the number of publications dealing with plasma-activated liquids (PAL) has dramatically increased. If PALs are widely used in the biomedical field with applications in the treatment of cancer (Tanaka et al., 2016; Boehm et al., 2016; Merbahi et al., 2017), inactivation of bacteria (Zhang et al., 2016; Shen et al., 2016) or fungi (Panngom et al., 2014), they are still poorly investigated in agriculture applications. Still, the potential of CAP seems tremendous in regard of emerging works performed on various agronomic models (lentils, radishes, tomatoes and sweet peppers) and showing important effects on seeds germination promotion and seedlings stems elongations (Zhang et al., 2017; Sivachandiran and Khacef, 2017; Lindsay et al., 2014). In that framework, CAP activations could find several additional benefits to boost crop yields: first as a cleaning process of water allowing the reduction of micro pollutants and microorganisms and second as a process permitting the synthesis of radicals-based green fertilizers. The objective of this work is to precisely evaluate the chemical interactions between cold atmospheric plasmas and tap water to identify the long lifetime species that could have a beneficial impact in the treatment of seeds.

In this study, we present a simple set of techniques enabling the quantification of 11 chemical species by colorimetric method (ammonia, ammonium, orthophosphates, nitrites, nitrates and hydrogen peroxide), 3 species by acid titration (carbonate ions) and 2 species by temperature and pH measurements. The predominance of the acid-base pairs is evaluated considering selective protocols and expression of Ka equilibria constants. All the concentrations measurements include uncertainties calculations taking into account variations in pH and temperature resulting from the liquid activation. Chemical reactions leading to variations in concentrations of these chemical species are discussed as well as the influence of plasma activation time and configuration of the DBD reactor (with/without hermetic enclosure).

A chemical model of plasma-activated Tap water "PATW" electrical conductivity ($\sigma_{PATW}$) has been developed according to the Kohlrausch's law: each individual conductivity of the ions present in the PATW is calculated, their sum leading to the theoretical $\sigma_{PATW}$. The values predicted by our model are compared with experimental measurement of $\sigma_{PATW}$. The appropriate fitting between simulation and experimental curves clearly demonstrate the self-sufficiency of our experimental approach, meaning that the experimental set includes the necessary techniques for the characterization of all relevant long lifetime chemical species in PATWs.

Finally, to demonstrate the potential of plasma-activated water in life sciences applications and in particular in agriculture, PATW has been utilized as an irrigation liquid for coral lentils (lens culinaris). This agronomical model has already been used by our team to investigate biological phenomena such as seeds dormacy, germination and early-stages of seedlings development (Zhang et al., 2017). They present a very low dormancy and a high germination rate. Results dealing with germination and stems elongation are presented.

## 2. Experimental setup

### 2.1. Plasma source & diagnostics of the plasma phase

Water has been activated using a dielectric barrier discharge (DBD) in a configuration dedicated to agronomical applications. An AC voltage delivered by a function generator (ELC Annecy France, GF467AF) is augmented by a power amplifier (Crest Audio, 5500W, CC5500) and supplies this DBD at 500 Hz. A ballast resistor (250 kΩ, 600W) protects the power supply from any excessive currents, as shown in Fig. 1 b. The excitation electrode is composed of a metal plate from which a matrix of 4*4 metal rods are embedded in dielectric tubes, placed in front of the tap water to be treated. In the entire study, the voltage applied on this electrode has an amplitude of 12.0 $kV_{AC}$ at 500 Hz. A volume of 50 mL of tap water is immerged into a Petri dish (diameter 136 mm) and directly connected to the ground electrode. A gap distance of 4mm is observed between the liquid interface and each dielectric tube containing a rod electrode. Also, the water layer thickness of 3.44mm is kept constant throughout the treatment (±5%) by an external regulatory system. The experimental setup shown in Fig. 1 has been used in two configurations: operation in ambient air and operation in a small hermetic enclosure so as to accumulate in a small volume the chemical radicals resulting from gas conversion and force their interaction with the water contained in the glass vessel.

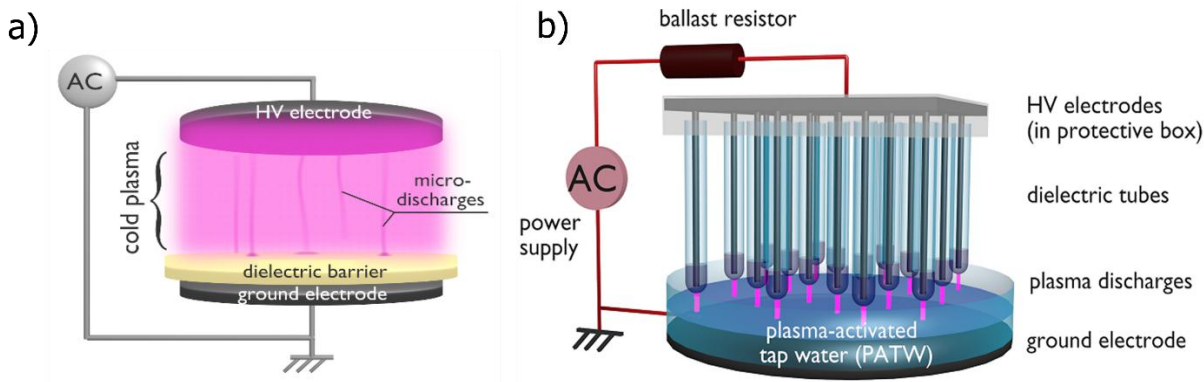

Fig. 1. Basic diagram of DBD plasma generation (a) and experimental setup of DBD plasma for activation of tap water (b).

## 2.2. Water & related diagnostics

### 2.2.1. pH and temperature

pH and temperature are both monitored using the PS-2147 probe from Pasco Scientific. The temperature can be measured between – 10°C and +70°C with an accuracy of 0.5°C while the pH between 0 and 14 with an accuracy as high as 0.1. The probe is connected to a computer using a high-resolution amplifier and a communication module (PS 3200, Pasco Scientific). Real-time tracking of sensor responses is performed with Pasco Capstone software (Version: 1.7.0, Pasco Scientific). The temperature measurement is carried out by fast response thermocouple in a volume of 50 mL of tap water while it is exposed to the plasma to prevent any cooling effect. During all treatments, the temperature outside the hermetic enclosure is thermostated at 27°C by an air conditioning/heating system.

### 2.2.2. Conductivity

Electrical conductivity ($\sigma$) of tap water is evaluated with a conductometer (HI-87314, Hannah) on the 199.9 µS/cm - 199.9 mS/cm range with an accuracy of 1% (full scale). This device is calibrated before any measurements using a dedicated standard solution ($\sigma$ = 1413 µS/cm) at 25°C. Also, since electrical conductivity depends on the liquid temperature, the conductometer includes a linear algorithm that displays s measurements for a reference temperature of 25°C.

### 2.2.3. Ammonia and ammonium ions ($NH_3$, $NH_4^+$)

Ammonia and ammonium ions are quantified using Nessler reagent. This one reacts with ammonia in basic solution to form an orange color complex with a specific absorbance at 400 nm observed by spectrophotometer (7315, Jenway). 2 mL of this reagent is added to 50 mL of PATW, and its absorbance spectrum is plotted 10 min after. To prevent any thermal side effect on quantification, the plasma-activated water samples are all stored at room temperature before mixing with Nessler reagent. Calibration is obtained from ammonium chloride to provide linear relation between absorbance at 400 nm and ammonia concentration from 0 to 277 µmol/L.

### 2.2.4. Hydrogen peroxide ($H_2O_2$)

To convert hydrogen peroxide into a species photometrically detectable, a reagent such as Fe (II), Co (II) or Ti (IV) compound (e.g. titanium oxysufate, titanium chlorate) is required (Pashkova et al., 2009). Here, an aqueous and acidic solution of titanium oxysulfate from Sigma Aldrich is purchased. It reacts in presence of $H_2O_2$ to produce a yellow peroxotitanium complex $[Ti(O_2)OH(H_2O)_3]^+_{aq}$ presenting an absorbance peak at 409 nm (Reis et al., 1996):

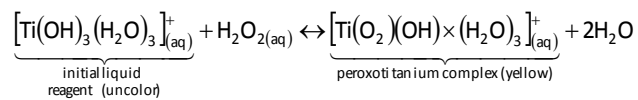

$$\underbrace{\left[Ti(OH)_3(H_2O)_3\right]^+_{(aq)}}_{\text{initial liquid reagent (uncolor)}} + H_2O_{2(aq)} \leftrightarrow \underbrace{\left[Ti(O_2)(OH)\times(H_2O)_3\right]^+_{(aq)}}_{\text{peroxotitanium complex (yellow)}} + 2H_2O$$

First, the $TiOSO_4$ probe is calibrated using different volumes of commercial $H_2O_2$ solution (30 wt % from Sigma Aldrich, CAS 7722-84-1). Then, 5mL of $TiSO_4$ reagent is mixed with 2mL of PATW to measure its $H_2O_2$ concentration. In accordance with the Beer-Lambert law, the absorbance measured at 409 nm is proportional to the $H_2O_2$ concentration, giving a linear relationship from 0 to 5 mmol/L.

### 2.2.5. Orthophosphates ($H_3PO_4$, $H_2PO_4^-$, $HPO_4^{2-}$, $PO_4^{3-}$)

Concentrations of orthophosphates are evaluated in PATW using the ascorbic acid method, which allows measurements in the range 0-32 µmol/L. In an acid medium, orthophosphate reacts with ammonium molybdate and antimony potassium tartrate to form phosphomolybdic acid. This heteropoly acid is then reduced by ascorbic acid to an intensely blue-colored complex presenting a specific absorbance at 720 nm. During this procedure, 5 mL of PATW is mixed with reagent (composed of ammonium molybdate [6 g.L$^{-1}$], antimony potassium tartrate [145 mg.L$^{-1}$], sulfuric [227 g.L$^{-1}$] and ascorbic acid [20 g.L$^{-1}$]). Owing to the instability of the complex, the analysis is performed 15 min after adding reagent. Calibration curve is obtained from dilute solution of potassium phosphate monobasic (Sigma Aldrich, P5379).

### 2.2.6. Carbonate ion ($CO_3^{2-}$), bicarbonate ($HCO_3^-$) and carbonic acid ($H_2CO_3$)

$CO_3^{2-}$, $HCO_3^-$ and $H_2CO_3$ concentration in PATW are assayed with acid titration. A titration curve is obtained by adding the HCl titrant ($V_t$, $C_t$=26.37 mmol/L) to the PATW analyt ($V_a$=50 mL) while monitoring pH. Then, the equivalence point is estimated from first derivate of titration curve. At equivalence point, concentration of analyte ($C_a$) is estimated through the following formula:

$$C_a = \frac{C_t \times V_t}{V_a}$$

### 2.2.7. Nitrite ($NO_2^-$), nitrous acid ($HNO_2$), nitrate ($NO_3^-$), Nitric acid ($HNO_3$)

Nitrites are quantified by colorimetric assay using Griess reagent (Sigma-Aldrich Co., Ltd, CAS Number 1465-25-4). Griess reagent is composed of two main reagents. The first is sulphanilic acid with which nitrites react to form diazonium salt. The second is naphthylethylenediamine dihydrochloride and is used to form azo dye agent with diazonium salt. Azo dye agent develops a pink color with typical absorbance at 540 nm (spectrophotometer 7315, Jenway). Before observation with spectrophotometer, Griess reagent is mixed in 1:1 vol with PATW (i.e. 1.5mL of Griess reagent with 1.5mL of PATW). As calibration curve is plotted from initial solutions of $NaNO_2$ (Sigma-Aldrich Co., Ltd.) in a range of concentrations between 0 and 50 µmol/L of $NO_2^-$, PATW is diluted 1/10 before each measurement to have adequate concentration. Formation of nitrate in PATW is investigated using $NO_3^-$ probe (CI-6735, Pasco Scientific). A calibration curve is plotted from initial solutions of $NO_3^-$ (0.1M nitrate standard, Pasco Scientific) in a range of concentrations between 100 µmol/L and 11.7 mmol/L. Before each measurement, 1mL of nitrate ISA (nitrate ISA, 2M $(NH_4)_2SO_4$, Pasco Scientific) is added into 50 mL of PAW. Finally, nitric and nitrous acids concentrations are obtained from respectively nitrate and nitrite concentration using acid-base equilibria.

### 2.3. Seeds, environment & diagnostics

In the present work, we have evaluated seeds irrigation considering two conditions: untreated tap water (UTW) and plasma-activated tap water (PATW). For each condition, one hundred seeds are placed into a germinator consisting of a small glass jar including lid with drainage holes. The protocol is quite simple and is declined in two phases: the first day, seeds are totally immersed 3h in 50 mL of UTW or PATW treated for 15 min (PATW$_{15}$). The days after, seeds are irrigated twice a day with 50 mL of UTW or PATW$_{15}$ during 3 min, the excess of water being removed from germinators just after. The experiments are performed at room temperature (22.5±0.5°C) with relative humidity ranging between 35 and 45%. Seeds are stored into darkness during all their dormancy phase. Once they have germinated, they are exposed 16 h/day to horticultural dimmable Led panel mimicking solar irradiation. The emission spectrum of this lamp is described in previous paper (Zhang et al., 2017). Finally, no soil or any organic substrate is utilized during experiments to clearly demonstrate the impact of PATW on seeds and seedlings biology without any plausible interference of soils or growing substrates properties.

# 3. Results

## 3.1. Characterization of the liquid phase

### 3.1.1. Temperature

Upon its plasma activation, tap water is gradually heated, inducing an increase in its temperature as reported in Fig. 2. Starting at ambient room temperature (27°C), the tap water under plasma exposure shows an increase of 3.60 ± 0.72°C after 5 min and 9.03 ± 0.78°C after 30 min, which corresponds to an absolute temperature of 36.15 ± 1.26°C. The rise in temperature is not linear: after 25 min of activation (90% response time), water heating slows down and its temperature reaches a plateau estimated at 36.73 ± 1.26°C. A nonlinear regression of water temperature variation gives the following equation ($R^2$=0.96):

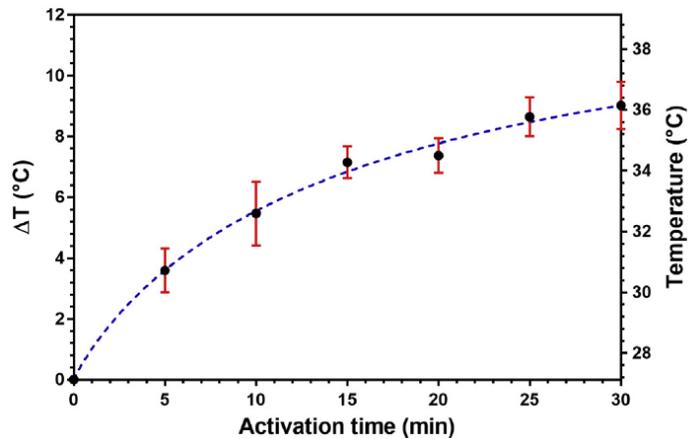

$$\Delta T = \frac{14.05 \times t_a^{0.92}}{t_a^{0.92} + 15.88^{0.92}}$$

*Fig. 2. Temperature of plasma-activated Tap water for different DBD-plasma treatment times.*

In this equation:

- ΔT is the variation of water temperature after plasma treatment in °C

- $t_a$ is the activation time in minutes

- 14.05°C is maximal variation of water temperature using this plasma treatment

- 15.88 min is the time activation to reaches 50% of maximal variation temperature

This maximal variation of water temperature validates the good agreement between our plasma source and agriculture applications, e.g. indirect treatment of seeds. Beyond 30 min, the plasma device seems always compatible with processing of heat-sensitive equipment.

The temperature plateau that could be reached for higher activation times illustrates a thermal equilibrium with the plasma source. Since water is not an adiabatic system here, this equilibrium illustrates that there is as much heat received from the plasma as heat given to the ambient air. Gas temperature of the plasma may be evaluated using optical emission spectroscopy on hydroxide and molecular nitrogen systems, in a temperature range between 300 K and 350 K.

### 3.1.2. pH

Since pH measurements are performed immediately after plasma activation, they are distorted by the higher temperature of the liquid which can lead to (i) a decrease in its viscosity and therefore an increase in its ion mobility and (ii) a higher dissociation of molecules inducing an increase in its overall ion concentration (Barron et al., 2011). A re-calibration process is therefore a mandatory and has been achieved using the Nernst's equation:

$$pH = \frac{E_0 - E}{2.3 \frac{R \times T}{F}}$$

Where F is the Faraday constant and R the universal gas constant. Standard potential ($E_0$) and 2.3R/F coefficient is estimated with two point calibration in buffer solutions of pH 4 and 7.

pH variations during the plasma activation is plotted in Fig. 3 where a linear regression clearly indicates only a negligible decay of the pH, from 7.8 (0 min) to 7.6 (30 min). pH standard deviation can be considered as negligible relative to other pH measurements after plasma activation (except for activation time of 25 min). It turns out that overall pH variations are negligible, making this parameter as a non-disruptive factor of the germination process.

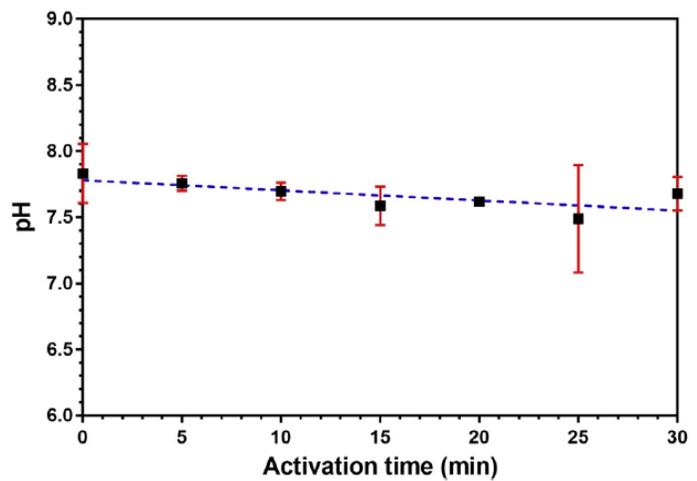

*Fig. 3. Variation of pH in Tap water after plasma treatment at various exposure time.*

Hydroxide ions concentration is calculated from both pH and ionic product of PATW. Ionic product of water, also called self ionization constant of water, represents the self-dissociation of water molecules into hydroxide ions and hydronium ions and is given by $K_W=[H_3O^+].[OH^-]$. $K_W$ depends on temperature, pressure and ionic strength. Since in our experiments only variations in temperature are observed, $pK_W$ can be estimated with the following formula (data extract from (Harned and Owen, 1958)):

$$pK_W = 14:88 - 0.0335.T$$

Hydroxide ions concentration and its standard deviation are finally determined combining the two mentioned equations, leading to the following formula:

$$[OH^-] = 10^{pH + 0.0335 \times T - 14.88} \quad (R1)$$

$$\Delta[OH^-] = \left|0.0335 \times \ln(10) \times 10^{pH + 0.0335 \times T - 14.88}\right|\Delta T + \left|\ln(10) \times 10^{pH + 0.0335 \times T - 14.88}\right|\Delta pH$$

Unsurprisingly, hydroxide concentration is not impacted by plasma treatment, showing an average value close to 0.75 µmol/L (data not shown). This stability versus time may result from a compensation of two simultaneous effects: a decrease in pH and an increase in temperature (see equation (R1)).

### 3.1.3. Electrical conductivity

The electrical conductivity can provide information on both nature and concentrations of ions present in electrolytic solution. Electrical conductivity of PATW is shown in Fig. 4. During the first 5 min of activation, σ decreases from 647.33 ± 15.07 µS/cm to 614 ± 10.60 µS/cm. Although not plotted in the figure, a minimum value of conductivity is usually observed between 0 and 5 min of plasma activation. For plasma activation times higher than 5 min, electrical conductivity increases linearly with the activation time, on a range covering values from 614 ± 10.60 µS/cm to 731.33 ± 19.37 µS/cm for 5 min and 30 min respectively. Linear regression of this part gives following equation ($R^2$=0.99):

$$\sigma = 591.20 + 4.74 \times t_a$$

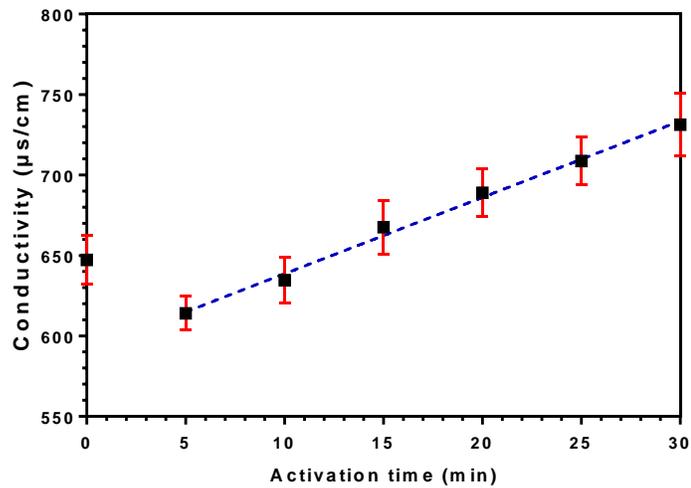

*Fig. 4: Electrical conductivity variation of tap water after plasma treatment*

### 3.1.4. Ammonia [$NH_3$] and ammonium ions [$NH_4^+$]

If ammonium ($NH_4^+$) and nitrate are the most important source of inorganic nitrogen for plants (Jampeetong and Brix, 2009), many studies show that too high concentrations lead to growth inhibition (Jampeetong and Brix, 2009; Roosta et al., 2009). For example ammonium concentration higher than 0.5 mM usually induces significant toxicity.

In water, $NH_4^+$ and $NH_3$ form a conjugate acid-base pair whose dissociation constant potential is $pK_a$ = 9.25. Dissociation constant is a specific type of equilibrium constant that measures the value of pH required to obtain rigorously same concentration values of acid and base in solution. Therefore, the base is dominant if pH > $pK_a$ and conversely. Since in our PATW, pH = 7.6-7.8, it is lower than $pK_a(NH_4^+/NH_3)$, meaning that ammonium ($NH_4^+$) – the acid component – is dominant. Concentration of each species from acid-base pair can be obtained from the following equation:

$$pH = pK_a + \log\frac{[Base]}{[Acid]} = pK_a + \log\frac{[NH_3]}{[NH_4^+]}$$

Nessler reagent is commonly used for quantification of ammonia in basic solution. Concentration of species provided by Nessler reagent [Total] is composed in our experiment of ammonia present in PATW but also of ammonium ions. Using the preceding formula, $NH_3$ and $NH_4^+$ concentrations as well as standard deviations are evaluated using following equations:

| Concentrations | Uncertainties |
|---|---|
| $[NH_4^+] = \dfrac{[Total]}{1 + 10^{pH-pKa}}$ | $\Delta[NH_4^+] = \left\|\dfrac{1}{1 + 10^{pH-pKa}}\right\|\Delta[Total] + \left\|\dfrac{-[Total] \times \ln(10) \times 10^{pH-pKa}}{(1 + 10^{pH-pKa})^2}\right\|\Delta pH$ |
| $[NH_3] = \dfrac{[Total] \times 10^{pH-pKa}}{1 + 10^{pH-pKa}}$ | $\Delta[NH_3] = \left\|\dfrac{10^{pH-pKa}}{1 + 10^{pH-pKa}}\right\|\Delta[Total] + \left\|\dfrac{[Total] \times \ln(10) \times 10^{pH-pKa}}{(1 + 10^{pH-pKa})^2}\right\|\Delta pH$ |

Fig. 5 shows that both ammonia and ammonium ions increase linearly with the activation time but to different extends respectively $R^2$=0.86 and $R^2$=0.99. Concentration of ammonia slightly increases with activation time from 0.42±0.25 µmol/L for UTW to 2.12 ± 0.31 µmol/L for an activation of 30 min. On the contrary $NH_4^+$ concentration quickly increases from 9.88 ± 3.40 µmol/L to 70.25 ± 0.75 for 30 min. Ammonia and ammonium ions concentration can be estimated using following linear regressions:

$$[NH_4^+] = 2.05 \times t_a + 10.15$$

$$[NH_3] = 0.05 \times T_a + 0.42$$

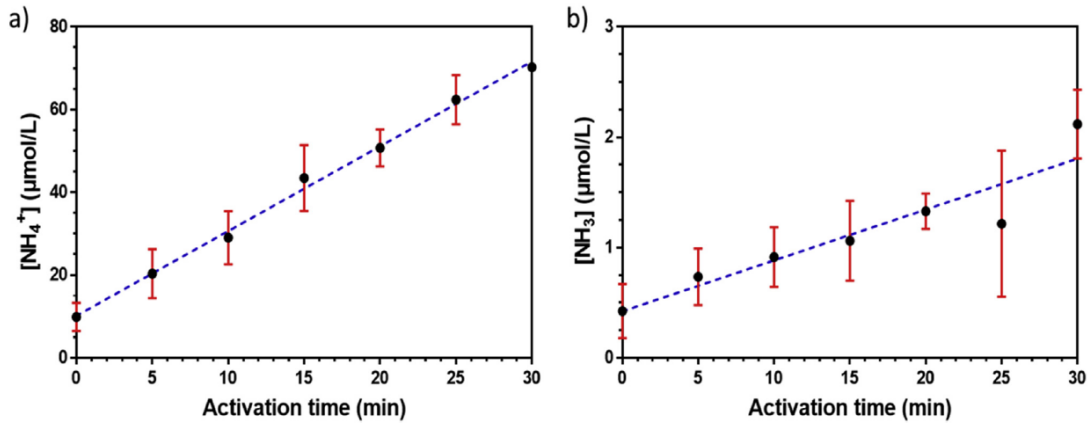

*Fig. 5: Linear increase of ammonium ions (a) and ammonia (b) concentration in Tap water after plasma treatment.*

### 3.1.5. Hydrogen peroxide ($H_2O_2$)

Hydrogen peroxide is a reactive oxygen species which also plays the role of a signaling molecule in redox biology and therefore in life applications, e.g. medicine and agriculture. In medicine, hydrogen peroxide plays a role in immune system (Niethammer et al., 2009). Even if ROS have a paradoxical effect on cancer cell proliferation (Vilema-Enríquez et al., 2016). Recently some publications have highlighted plasma-induced hydrogen peroxide induced by plasma as a new approach of cancer treatment (Boehm and Bourk, 2017; Judée et al., 2016; Kaushik et al., 2015).

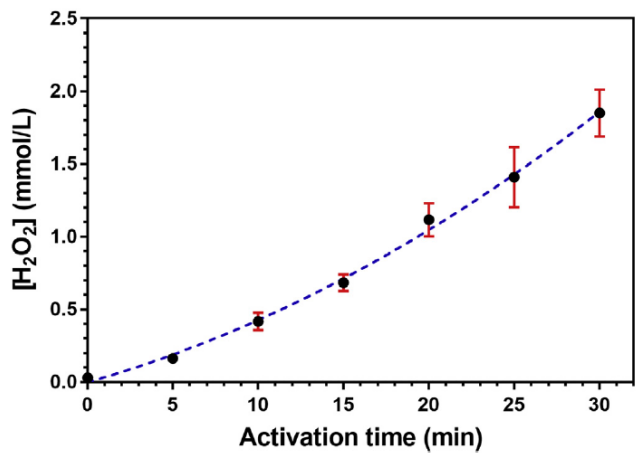

In agriculture, hydrogen peroxide can participate to the dormancy release by down-regulating the blockage of abscisic acid (ABA) and induce great effects on development, fruit growth and quality (Ismael et al., 2015).

Hydrogen peroxide in PATW is quantified using a solution of titanium oxysulfate and data are shown in Fig. 6. Plasma activation generates high concentration of $H_2O_2$ in tap water with a maximum value of 1.85 ± 0.16 mmol/L for 30 min. Experiments indicate quadratic relation between hydrogen peroxide concentration and activation time ($R^2$ = 0.94) according formula:

$$[H_2O_2] = 0.039 \times t_a + 0.00096 \times t_a^2$$

*Fig. 6. Hydrogen peroxide concentration measurements during plasma treatment.*

### 3.1.6. Phosphate ($PO_4^{3-}$), hydrogen phosphate ($HPO_4^{2-}$), dihydrogen phosphate ($H_2PO_4^-$) and phosphoric acid ($H_3PO_4$)

The ascorbic acid method is used to determine any variation of orthophosphates in PATW. Whatever the activation time, the orthophosphate concentration remains low and stable with an average value estimated at 10.84 ± 0.97 µmol/L (data not shown). From this orthophosphate concentration and pH measurements, all the related concentration in phosphates can be determined using dissociation constant of each acid/base couple given in Table 1 with $A = 10^{pH-pKa_1}$, $B = 10^{-pH+pKa_2}$, $C = 10^{-2.pH+pKa_2+pKa_3}$ and $D = 10^{2.pH-pKa_2-pKa_3}$. These formulas allow calculating concentration values and uncertainties of phosphate, hydrogen phosphate, dihydrogen phosphate and phosphoric acid. The data are reported in Table 2. As pH value decreases slowly with activation time, concentration of phosphate species decreases for base with $pK_a > pH$ (phosphate and hydrogen phosphate) and increase for acid with $pK_a < pH$ (dihydrogen phosphate and phosphoric acid). Before and after plasma activation of water, only hydrogen and dihydrogen phosphates show significant concentrations, in the order of several µmol/L.

In the case of phosphoric acid and phosphate, the standard error of the mean (SEM) values appear quite elevated owing to their low concentrations combined with significant uncertainties of the pH values.

| Acid | Base | $pK_a$ |
|---|---|---|
| $H_3PO_4$ | $H_2PO_4^-$ | 12.42 |
| $H_2PO_4^-$ | $HPO_4^{2-}$ | 7.2 |
| $HPO_4^{2-}$ | $PO_4^{3-}$ | 2.15 |
| Concentrations | Uncertainties | |
| $[HPO_4^{2-}] = \frac{[Total]}{1+A+B+C}$ | $\Delta[HPO_4^{2-}] = \left\vert \frac{1}{1+A+B+C} \right\vert \Delta[Total] + \left\vert \frac{-[Total]\times[(A-B-2C).\ln(10)]}{(1+A+B+C)^2} \right\vert \cdot \Delta pH$ | |
| $[PO_4^{3-}] = 10^{pH-pKa_1} \times [HPO_4^{2-}]$ | $\Delta[PO_4^{3-}] = \left\vert 10^{pH-pKa_1} \right\vert \Delta[HPO_4^{2-}] + \left\vert [HPO_4^{2-}] \times \ln(10) \times 10^{pH-pKa_1} \right\vert \Delta pH$ | |
| $[H_2PO_4^-] = \frac{[HPO_4^{2-}]}{10^{pH-pKa_2}}$ | $\Delta[H_2PO_4^-] = \left\vert \frac{1}{10^{pH-pKa_2}} \right\vert \Delta[HPO_4^{2-}] + \left\vert \frac{-[HPO_4^{2-}]\times \ln(10) \times 10^{pH-pKa_2}}{(10^{pH-pKa_2})^2} \right\vert \Delta pH$ | |
| $[H_3PO_4] = \frac{[HPO_4^{2-}]}{10^{2\times pH-pKa_2-pKa_3}}$ | $\Delta[H_3PO_4] = \left\vert \frac{1}{D} \right\vert \Delta[HPO_4^{2-}] + \left\vert \frac{-[HPO_4^{2-}]\times 2 \times \ln(10) \times D}{D^2} \right\vert \Delta pH$ | |

*Table 1. Value of acidity constant for phosphates acid/base couple.*

| | | Activation time (min) | | | | | | |
|---|---|---|---|---|---|---|---|---|
| | | 0 | 5 | 10 | 15 | 20 | 25 | 30 |
| $[H_3PO_4]$ (pM) | Mean | 4.52 | 5.8 | 8.11 | 11.15 | 9.76 | 16.20 | 7.55 |
| | SEM | 2.89 | 1.6 | 1.65 | 5.09 | 2.11 | 20.20 | 4.29 |
| $[H_2PO_4^-]$ (µM) | Mean | 2.18 | 2.35 | 2.86 | 3.05 | 2.87 | 3.56 | 2.56 |
| | SEM | 0.83 | 0.48 | 0.36 | 0.88 | 0.52 | 2.75 | 1.08 |
| $[HPO_4^{2-}]$ (µM) | Mean | 9.34 | 8.48 | 9.00 | 7.45 | 7.52 | 6.92 | 7.71 |
| | SEM | 1.15 | 1.21 | 0.42 | 0.92 | 1.12 | 2.11 | 2.14 |
| $[PO_4^{3-}]$ (pM) | Mean | 242 | 184 | 171 | 110 | 119 | 91 | 140 |
| | SEM | 92 | 38 | 21 | 32 | 22 | 63 | 59 |

*Table 2. Phosphate, hydrogen phosphate, dihydrogen phosphate and phosphoric acid in (PA)TW: concentrations and SEM values.*

### 3.1.7. Carbonate ion ($CO_3^{2-}$), bicarbonate ($HCO_3^-$) and carbonic acid ($H_2CO_3$)

Carbonate ion, bicarbonate and carbonic acid are quantified by acid titration using HCl as titrant. The addition of this strong acid into water generates hydronium ions ($H_3O^+$) that can react with the species contained in the PATW, according to the following reactions (see on the right).

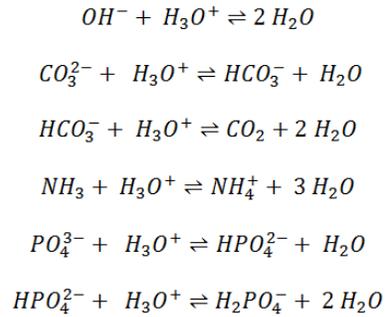

$$OH^- + H_3O^+ \rightleftharpoons 2\,H_2O$$
$$CO_3^{2-} + H_3O^+ \rightleftharpoons HCO_3^- + H_2O$$
$$HCO_3^- + H_3O^+ \rightleftharpoons CO_2 + 2\,H_2O$$
$$NH_3 + H_3O^+ \rightleftharpoons NH_4^+ + 3\,H_2O$$
$$PO_4^{3-} + H_3O^+ \rightleftharpoons HPO_4^{2-} + H_2O$$
$$HPO_4^{2-} + H_3O^+ \rightleftharpoons H_2PO_4^- + 2\,H_2O$$

Consequently, volumetric analysis provides information on concentration of carbonate ion and bicarbonate but also on hydroxide ion, ammonia and phosphates. Concentration and uncertainties of $CO_3^{2-}$, $HCO_3^-$ and $H_2CO_3$ can be calculated from equivalent point with following formulas and Table 3:

$$V_{solution} \times ([OH^-] + 2[CO_3^{2-}] + [HCO_3^-] + [NH_3] + 2[PO_4^{3-}] + [HPO_4^{2-}]) = V_{equiv} \times [HCl]$$

$$V_{solution} \times (2[CO_3^{2-}] + [HCO_3^-] + \xi) = V_{equiv} \times [HCl] \quad \text{with} \quad \xi = [OH^-] + [NH_3] + 2[PO_4^{3-}] + [HPO_4^{2-}]$$

| Acid | Base | pK$_a$ | pH |
|---|---|---|---|
| $HCO_3^-$ | $CO_3^{2-}$ | 10.33 | $pH = pKa1 + \log\frac{[CO_3^{2-}]}{[HCO_3^-]}$ |
| $H_2CO_3$ | $HCO_3^-$ | 6.3 | $pH = pKa2 + \log\frac{[HCO_3^-]}{[H_2CO_3]}$ |

| Concentrations | Uncertainties |
|---|---|
| $[Total] = 2[CO_3^{2-}] + [HCO_3^-]$   $[Total] = \frac{V_{equiv}}{V_{solution}} \times [HCl] - \xi$ | $\Delta[Total] = \left\|\frac{V_{equiv}}{V_{solution}}\right\|\Delta[HCl] + \left\|\frac{[HCl]}{V_{solution}}\right\|\Delta V_{equiv} + \Delta[OH^-] + \Delta[NH_3] + 2\Delta[PO_4^{3-}] + \Delta[HPO_4^{2-}]$ |
| $[CO_3^{2-}] = \frac{[Total] \times 10^{pH-pKa1}}{1+2\times 10^{pH-pKa1}}$ | $\Delta[CO_3^{2-}] = \left\|\frac{10^{pH-pKa1}}{1+2\times 10^{pH-pKa1}}\right\|\Delta[Total] + \left\|\frac{[Total]\times \ln(10)\times 10^{pH-pKa1}}{(1+2\times 10^{pH-pKa1})^2}\right\|\Delta pH$ |
| $[HCO_3^-] = [Total] - 2[CO_3^{2-}]$ | $\Delta[HCO_3^-] = \Delta[Total] + 2 \times \Delta[CO_3^{2-}]$ |
| $[H_2CO_3] = \frac{[HCO_3^-]}{10^{pH-pKa2}}$ | $\Delta[H_2CO_3] = \left\|\frac{1}{10^{pH-pKa2}}\right\|\Delta[HCO_3^-] + \left\|\frac{-[HCO_3^-]\times \ln(10)\times 10^{pH-pKa2}}{(10^{pH-pKa2})^2}\right\|\Delta pH$ |

Table 3. Values of acidity constant potentials and pH for carbonates acid/base couple.

As shown in Fig. 7, bicarbonate decreases quickly with activation time, from 4.13 ± 0.09 mmol/L (untreated tap water) to 0.98 ± 0.25 mmol/L (plasma activation of 30 min). This decrease is non-linear, mainly observed in the first minutes of plasma activation. Bicarbonate concentration is fitted using dose-response inhibition formula ($R^2 = 0.95$):

$$[HCO_3^-] = -0.80 + \frac{4.94}{1+\frac{t_a^{0.73}}{11.73^{0.73}}}$$

Carbonate ions and carbonic acid are also quantified and resumed in Table 4. Relations with activation time are like the one obtained for bicarbonate, with a quick decrease in the first minute of plasma treatment. Unsurprisingly, concentrations of these species are lower than for bicarbonate, with 13.10 μmol/L and 121.33 μmol/L in untreated tap water or 2.20 μmol/L and 41.04 μmol/L for 30 min of activation respectively for carbonate ions and carbonic acid.

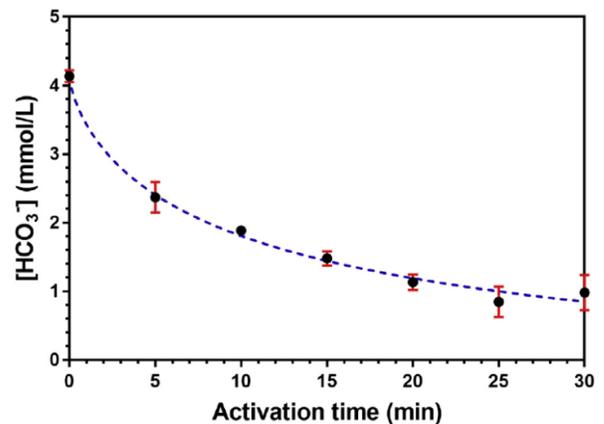

Fig. 7. Acid titration of bicarbonates ions in Tap water during plasma treatment.

|  |  | Activation time (min) | | | | | | |
|---|---|---|---|---|---|---|---|---|
|  |  | 0 | 5 | 10 | 15 | 20 | 25 | 30 |
| $[CO_3^{2-}]$ (µM) | Mean | 13.1 | 6.34 | 4.40 | 2.68 | 2.20 | 1.22 | 2.20 |
|  | SEM | 3.62 | 0.99 | 0.47 | 0.63 | 0.29 | 0.89 | 0.88 |
| $[H_2CO_3]$ (µM) | Mean | 121.33 | 82.63 | 75.50 | 76.33 | 54.31 | 54.76 | 41.04 |
|  | SEM | 30.83 | 13.07 | 8.09 | 17.95 | 7.34 | 40.00 | 16.59 |

*Table 4. Carbonate ions and phosphoric acid in (PA)TW: concentrations and SEM values*

### 3.1.8. Nitrite ($NO_2^-$), nitrous acid ($HNO_2$), nitrate ($NO_3^-$), Nitric acid ($HNO_3$)

Formation of nitrite in tap water during plasma activation is investigated using colorimetric assay. As shown in Fig. 8a, the concentration of nitrite in PATW increases with activation time during the first 10 min from 7.16 ± 0.23 µmol/L for UTW to 175.36 ± 9.67 µmol/L for 10 min of plasma activation. Then nitrite concentration is stable for activation time comprised between 10 and 15 min, before linearly decreasing for higher activation times, from 175.36 ± 9.54 µmol/L (15 min) to 125.79 ± 6.85 µmol/L (30 min). Nitrous acid is also detected using similar technique that for ammonia and ammonium but due to its very low $pK_a$ = 3.35, $[HNO_2]$ has a maximum value of 10.72 ± 5.78 nmol/L after 25 min of plasma activation and is considered here as negligible (data in Table 5).

The nitrate probe allows detection and quantification of $NO_3^-$ and $HNO_3$ in tap water after plasma exposure. Concentration of nitrite is plotted as a function of activation time in Fig. 8b while concentration in nitric acid are summarized in Table 5. Important concentration of nitrate is observed in UTW with value around 0.86 ± 0.09 mmol/L. This concentration increases near to linearly with activation time up to 3.55 ± 0.42 mmol/L after 30 min of plasma exposure. First derivate of nitrate concentration shows a decrease in the production rate with the exposure time (data not shown). As for nitrous acid, the $pK_a$ of nitric acid is too low (−2) to provide non-negligible concentration. Nitric acid concentration is estimated to be lower than 1.15 ± 0.69 pmol/L.

|  |  | Activation time (min) | | | | | | |
|---|---|---|---|---|---|---|---|---|
|  |  | 0 | 5 | 10 | 15 | 20 | 25 | 30 |
| $[HNO_2]$ (nM) | Mean | 0.24 | 4.73 | 7.88 | 10.15 | 8.60 | 10.72 | 5.89 |
|  | SEM | 0.07 | 0.50 | 1.05 | 2.24 | 0.55 | 5.78 | 1.18 |
| $[HNO_3]$ (pM) | Mean | 0.13 | 0.26 | 0.41 | 0.64 | 0.69 | 1.15 | 0.74 |
|  | SEM | 0.05 | 0.06 | 0.09 | 0.21 | 0.11 | 0.69 | 0.20 |

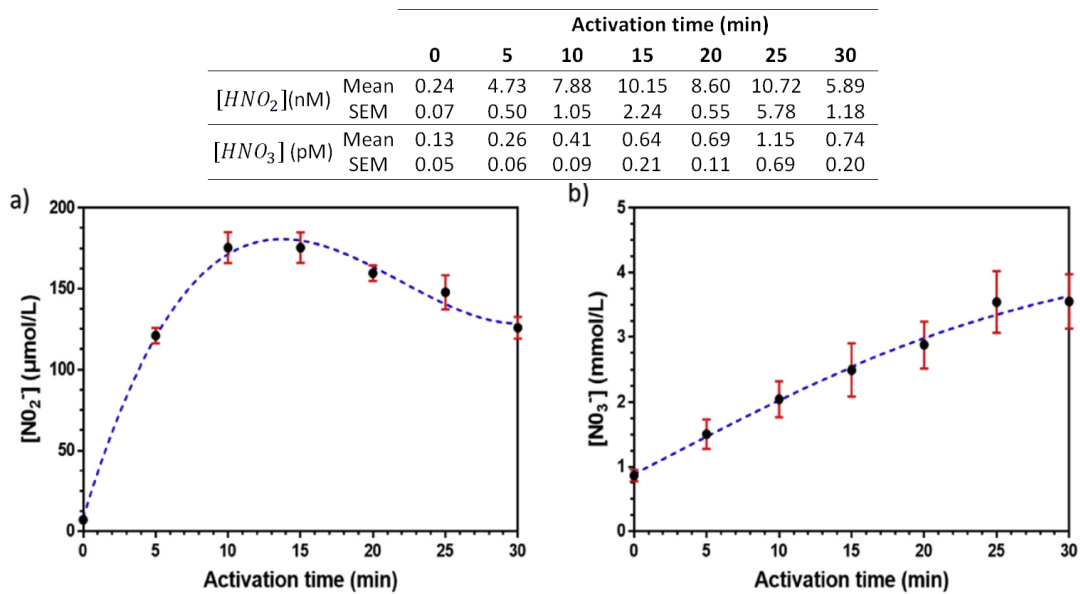

*Fig. 8. Variation of the concentration of (a) Nitrite anion and (b) Nitrate anion concentration in Tap water after DBD plasma treatment at various activation time*

## 3.2. Treatment of seeds with PATW

Fig. 9 a shows pictures of lentils seeds/seedlings taken at days 1, 3 and 6, considering untreated tap water and PATW. In both cases, the germination process is obtained in less than 24 h as claimed by the seed producer, with germination rates as high as 97% and 99% for UTW and PATW$_{15}$ respectively. Furthermore, temporal kinetics of seeds germination are similar regardless of liquid type: 50% of germination is reached in 11h16 (UTW) and 11h29 after the first treatment (data not shown). Two days after germination (Day 3), one can easily measure the total length of root and stem of each individual seedling. Mean length of lentils seeds treated with UTW is 1.4 ± 0.7 cm versus 1.97 ± 0.81 cm for PATW$_{15}$ which represents an increase of 38.0%. This trend is confirmed at day 6 with an increase in the total length of 128.41% for PATW$_{15}$ activation: 1.87 ± 0.95 cm (UTW) vs 4.28 ± 2.15 cm (PATW$_{15}$). Distribution lengths related to UTW and PATW activations are not identical, as reported in Fig. 11b at days 3 and 6. At day 3, both distributions show a gaussian shape while a peak-to-peak shift of 1 cm separates them, hence highlighting the plasma activation effect. At day 6, UTW distribution can always be fitted with a gaussian whereas distribution of PATW$_{15}$ is too random.

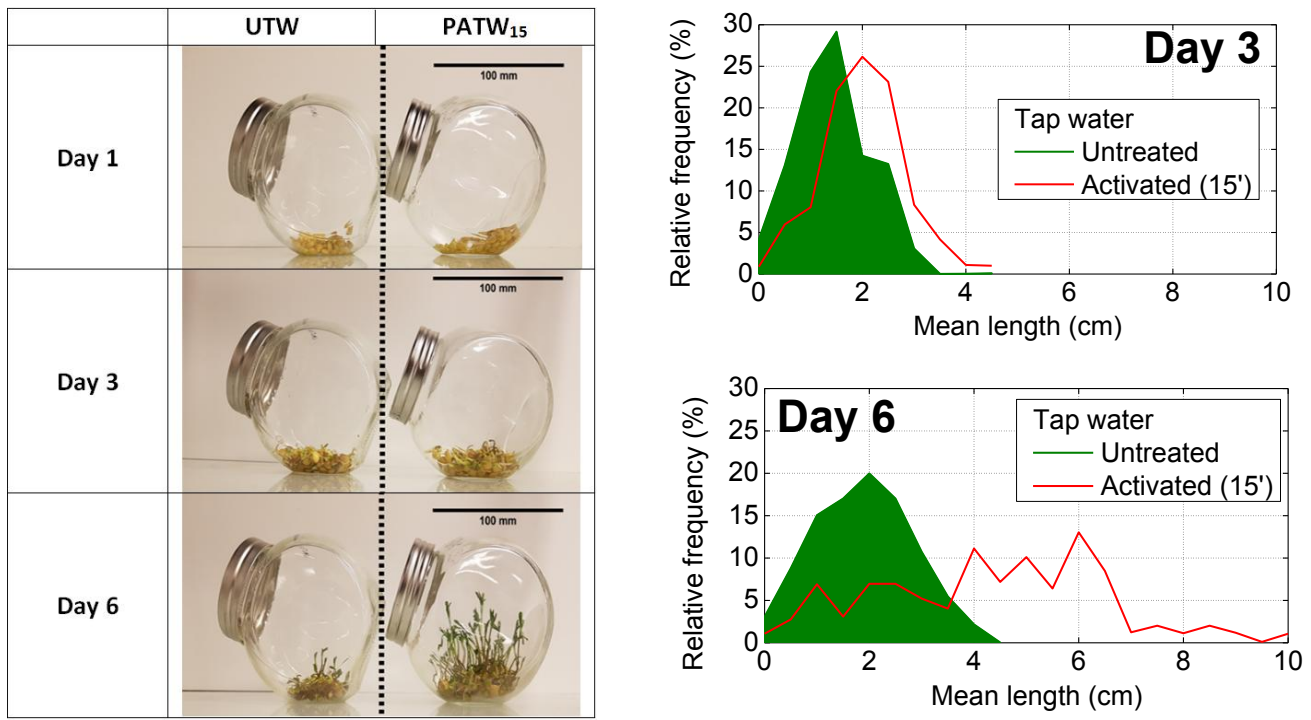

*Fig. 9. Effect of plasma-activated 15 min tap water (PATW$_{15}$) irrigation on coral lentils seeds: (a) pictures of plant growth 0, 3 and 6 days after the first irrigation with untreated Tap water (UTW) or PATW$_{15}$, (b) distribution of plants lengths 3 and 6 days after first treatment.*

# 4. Discussion

## 4.1. Electrical conductivity

Variations of $\sigma_{PATW}$ result from charged species production and consumption mechanisms occurring in the liquid phase. These mechanisms result from a complex interaction between the plasma and liquid phases. This interaction includes (i) the diffusion of gaseous species from the plasma to the liquid bulk, (ii) the stimulation of the liquid interface giving rise to new species in the subinterface layer and then to their in-depth diffusion and (iii) PATW heating effects induced by the plasma source.

Electrical conductivities of the PATWs have been estimated following two distinct approaches: an experimental approach where $\sigma_{PATW}$ is measured with the conductometer device and a theoretical approach based on the Kohlrausch's law. In low concentrated solutions (C <$10^{-2}$ mol.L$^{-1}$), the Kohlrausch's law establishes that the overall electrical conductivity is the sum of all the individual conductivities of the ions present in this solution (Prieve et al., 2017). Each individual ionic conductivity $\sigma_i$ can be estimated from its concentration $C_i$, its charge number $Z_i$ and its molar ionic conductivity at infinite dilution $\lambda_i$ (extract from (Lide, 2003)) as follows:

$$\sigma = \sum_{i=0}^{n} \sigma_i = \sum_{i=0}^{n} Z_i \times \lambda_i \times C_i$$

We have evaluated the PATW conductivity considering the 18 more influent ions. Their concentrations have been either measured upon this research work (hydroxide, hydronium, ammonium, carbonate, bicarbonate, nitrite, nitrate, orthophosphates, hydrogen phosphate, dihydrogen phosphate ions) or extracted from official data delivered by the Agence Régionale de Santé (Ile-de-France region) from the 5$^{th}$ district of Paris city (ARS, 2017) and summarized in Table 6.

| Ions | Concentration (µmol/L) |
|---|---|
| $Ca^{2+}$ | 2323 |
| $Cl^-$ | 640 |
| $Fe^{2+}$ | 0.0 |
| $F^-$ | 5.26 |
| $K^+$ | 56.3 |
| $Na^+$ | 392 |
| $SO_4^{2-}$ | 202 |
| $Mg^{2+}$ | 247 |

Table 6. Mean mineral salts composition in Tap water at Paris (ARS, 2017).

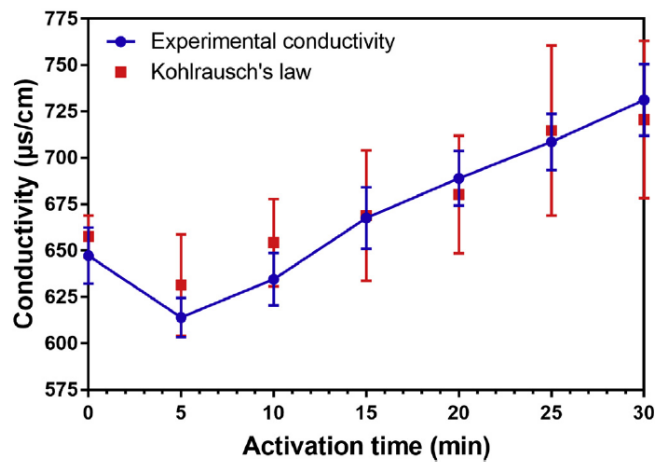

Fig. 10. Correlation between experimental electrical conductivity measurements and electrical conductivity calculated with ions concentration and Kohlraush's law.

As shown in Fig. 10, the electrical conductivities estimated by Kohlrausch's law are in good agreement with experimental measurements with a correlation factor r=0.98. Without any activation, $\sigma_{UTW}$ is mostly attributed to calcium ions (41.98%) and bicarbonate ions (27.96%). Then the part of ionic conductivity of bicarbonates ions in PATW electrical conductivity falls to 16.65% at 5 min and as low as 6.07% for 30 min of plasma activation. In the same time, the part of nitrates ionic conductivity rises linearly from 9.37% in UTW to 35.16% in $PATW_{30}$. Consequently, the fall of $\sigma_{PATW}$ during the first 5 min after plasma activation is imputable to the quick decrease of bicarbonates ions conductivity, i.e. their quick concentration decay. Conversely, all along the plasma activation, nitrates are generated in PATW and participate more and more to the global $\sigma_{PATW}$. This trend is clearly observed in Fig. 10: a linear increase of electrical conductivity is measured and calculated from 5 to 30 min of plasma activation.

## 4.2. NO$_X$ species

The atypical trend of nitrite concentration in water (already observed by Maheux et al. (2015)) during plasma activation is the result of a competition between nitrites and nitrates produced in a hermetic enclosure. The formation of nitrogen oxide in the gas phase is described by the Zeldovich's reactions (R2, R3, R4). In an air plasma, NO is rapidly converted to nitrogen dioxide by three body reactions with oxygen. In the second phase (R5, R6), nitrogen dioxide is dissolved in water, leading to the formation of nitrites, nitrates and hydronium ions.

|  |  |  |  |
|---|---|---|---|
| Phase 1 | $N_2 + O \rightarrow NO + N$ | (R2) | (Fridman, 2008) |
|  | $N + O_2 \rightarrow NO + O$ | (R3) | (Fridman, 2008) |
|  | $NO + O^\cdot + M \rightarrow NO_2 + M$ | (R4) | (Fridman, 2008) |
| Phase 2 | $2\,NO_{2\,(g)} + H_2O \rightarrow NO_{2\,(aq)}^- + NO_{3\,(aq)}^- + 2\,H_{(aq)}^+$ | (R5) | (Liu et al., 2016) |
|  | $NO_{(g)} + NO_{2\,(g)} + H_2O \rightarrow 2\,NO_{2\,(aq)}^- + 2\,H_{(aq)}^+$ | (R6) | (Vanreas et al., 2017) |

In open air, surrounded oxygen concentration is stable during plasma treatment, leading to a linear increase of nitrites and nitrates. In our case, the hermetic enclosure reduces the surrounded oxygen concentration with plasma exposure time due to Zeldovich's reaction for nitric oxide generation and ozone production.

$O_2^+$ + O + M → $O_3$ + M (Zhang et al., 2015; Van Gaens and Bogaerts, 2013)

Consequently, the nitrite production rate decreases with activation time while two reactions transform nitrites to nitrates using $O_3$.

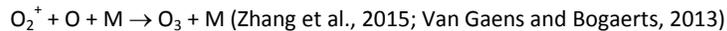

$NO_2^-$ + $O_3$ → $NO_3^-$ + $O_2$ (Liu et al., 2016; Van Gaens and Bogaerts, 2013)

Finally, the nitrite-to-nitrate conversion is reflected in nitrite concentration by a plateau comprised between 10 and 15 min and then a further decrease (see Fig. 8). Nitrites-nitrates competition mechanism explains also the slight decrease of pH during plasma activation, i.e. the formation of hydronium ions.

Ammonia is directly created in plasma phase due to interaction between excited nitrogen and hydrogen (Maheux et al., 2015). Once ammonia is solvated into water, it can be converted into ammonium ions by acid-base reaction.

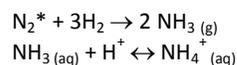

$N_2^* + 3H_2 \rightarrow 2\,NH_{3\,(g)}$
$NH_{3\,(aq)} + H^+ \leftrightarrow NH_{4\,(aq)}^+$

Linear increase of ammonium ions and ammonia in PATW induce that dihydrogen and excited molecular nitrogen present both stable concentrations in the gas phase during plasma activation.

### 4.3. Bicarbonates

The second species involved in $\sigma_{PATW}$ variations are bicarbonates ions. Their fast decrease upon plasma activation is not well understood. It is known that carbonate and bicarbonate ions are the representative species of gaseous carbon dioxide dissolved in water. Likewise, carbon dioxide dissolved in water decreases with a rise of water temperature. Therefore, a second assumption is that DBD activation heats very locally the plasma water at a temperature high enough to converse carbonate and bicarbonate ions into gaseous carbon dioxide. To verify that hypothesis, we have first correlated the variation of water temperature upon plasma activation with bicarbonate concentration. As the correlation coefficient is estimated to r = –0.985, one can reasonably validate this conjecture, i.e. a correlation between temperature and bicarbonate concentration. Gas analysis of carbon dioxide in activation chamber can be a good technique to validate the assumption of bicarbonate ions transformation into gaseous carbon dioxide. If this assumption is true, carbon dioxide concentration in activation chamber should increase with activation time.

### 4.4. Hydrogen peroxide

Aqueous and gaseous hydrogen peroxide can result from the recombination of two hydroxyl radicals in liquid or in plasma phase (Van Gaens and Bogaerts, 2013; Gorbanev et al., 2016). In plasma phase, hydrogen peroxide can be also generated from interaction between excited water molecule and hydroxyl radical (Maheux et al., 2015).

$$OH^\bullet + OH^\bullet \rightarrow H_2O_2$$
$$OH^\bullet + H_2O^* \rightarrow H_2O_2 + H^\bullet$$

Hydroxyl radical is formed in the plasma phase by the interaction of surrounded air (especially water molecules) with plasma through several reactions:

$$H_2O + e^- \rightarrow OH: þ H: þ e^- \text{ (Van Gaens and Bogaerts, 2013)}$$
$$H_2O^* + H_2O \rightarrow OH^\bullet + H^\bullet + H_2O \text{ (Zhang et al., 2015; Maheux et al., 2015)}$$
$$H_2O^* + H^\bullet \rightarrow OH^\bullet + H_2 \text{ (Maheux et al., 2015)}$$
$$H^\bullet + O \rightarrow OH^\bullet \text{ (Van Gaens and Bogaerts, 2013; Gorbanev et al., 2016)}$$

A specific signature of our DBD configuration is hydrogen peroxide kinetics during activation. The literature generally states a linear increase in $H_2O_2$ production with plasma exposure time (Chauvin et al., 2017), which is not the case here since the trend in Fig. 6 is clearly quadratic. In open air, surrounded water concentration is stable during plasma activation, leading to a linear increase of hydrogen peroxide. Here, when the DBD operates in a hermetic enclosure, water vapor concentration accumulates gradually with activation time: the relative humidity starts at 46.63 ± 0.80% (0 min) to reach a value as high as 89.33 ± 0.58% (30 min). The hermetic enclosure includes both the DBD plasma source and the glass vessel containing the liquid to be activated. From a thermodynamic point of view, this hermetic enclosure is not an isolated system but only a closed system, i.e. heat transfer with the ambient environment are possible but no matter transfer. Therefore water remains inside the hermetic enclosure, whatever its state: liquid or gaseous. Since water is activated by the DBD plasma device, its temperature is increased (+8°C in 25 min). Simultaneously, heating the water increases the saturation vapor pressure of water in the air contained in the hermetic enclosure. Higher concentration of water vapor increases the production rate of hydroxyl radicals in the plasma phase and therefore of hydrogen peroxide in PATW. The impact of relative humidity on hydrogen peroxide generation has already been observed in Gorbanev's works (Gorbanev et al., 2016) with DBD plasma jet using helium gas.

## 4.5. Increase of growth by plasma-activated tap water

Two observations on the present plasma-agronomy results are noteworthy. First, in our previous work where tap water was treated using a helium atmospheric pressure plasma jet, we obtained lentils stems elongations of 77% after 6 days (Zhang et al., 2017). In this study, the same lentils seedlings show stems elongation as high as 130% using a dedicated DBD as an alternative plasma source. Second, conversely to Sivachandiran's results on radish, tomato and sweet pepper, we do not observe any plasma effect on germination rate (Sivachandiran and Khacef, 2017).

As the seeds utilized in the present study are intended to consumers, they present a short native dormancy (24h) and a high germination rate (>95%). Therefore, the beneficial effects of PATW seem barely observable if (pre)germination kinetics are faster than signaling reactions. To clearly determine the effects of plasma activation on seeds dormancy release and germination rate, observations could be facilitated by slowing down biological processes, i.e. studying dormant seeds lots and sowing/irrigating them in a low temperature range, comprised between 10 and 15°C.

Since $H_3O^+$ and $OH^-$ concentrations are estimated close to 1 µmol, any species with concentrations lower than this threshold are considered negligible in this study and on the point of agronomical view. Therefore 8 chemical species could play a significant role on seeds germination and seedlings growth, as reported in Fig. 11a. Among them, $HPO_4^{2-}$ show no significant variation in plasma activation this species is discarded. Nitrates are present in PATW at 1-4 mmol/L while nitrites concentration increase from 7 to 180 µmol/L. According to Hendrick, nitrites and nitrates can promote *Amaranthus albus* seedlings growth, but also their germination for $[NO_3^-]$>1 mmol/L and $[NO_2^-]$>50 µmol/L (Hendricks and Taylorson, 1974). Conversely, concentrations exceeding 10 mmol/L for both species lead to a sharp decrease in the germination rate (Hendricks and Taylorson, 1974). More recently, the beneficial effect of nitrates on germination and growth has been observed for $[NO_3^-]$>100 µmol/L on *Arabidopsis* (Alboresi et al., 2005) and on *Paspalum vaginatum* (Shim et al., 2008). Since nitrite and nitrate ions generated in our PATWs show similar concentrations, they can reasonably affect agronomical models, including coral lentils for which a beneficial effect has been obtained.

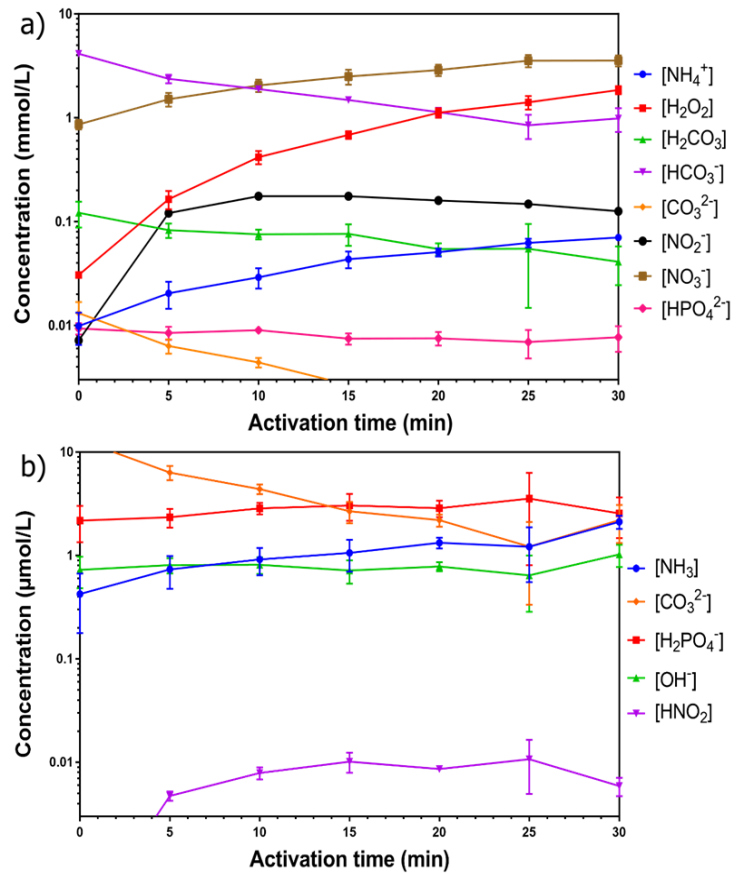

Fig. 11. Synthesis of major ion and neutral species concentration quantified in the plasma-activated Tap water for various activation time. (a) Only species concentration above 3 mmol/L are displayed and (b) only species concentration between 10 mmol/L and 3 nmol/L are displayed.

Indeed, hydrogen peroxide is known to be a signaling molecule with the ability to release dormancy (Ismael et al., 2015). It is therefore often cited as an important expedient for seed treatment. Moreover, it is also involved in ROS signaling and antimicrobial effect.

On cow pea seeds, concentration as low as 30 µM can result in dormancy release, higher germination rate, and increased seedling dry matter (Naim, 2015), although such positive effects seem limited to a maximum concentration of 20mM (Barba-Espín et al., 2012). In PATW$_{15}$, $H_2O_2$ concentrations is 0.7 mM, which makes this species an important candidate to support the higher growth of coral lentils.

A third group of species that can influence the growth of coral lentils include carbonates ions ($CO_3^{2-}$, $HCO_3^-$ and $H_2CO_3$). Indeed, in works on maize plants, it has been shown that increasing $HCO_3^-$ in water induces a decrease in length, dry mass and modifications in the concentration of cations and anions contained in plants (Massoud Al mansouri and Alhendawi, 2014). During the DBD activation, a significant decrease of bicarbonate ions concentration is observed, and could hence contribute to the higher growth observed in coral lentils after DBD activation.

In the present manuscript we hypothesized that plant growth gain was attributed to $H_2O_2$, $NO_2^-$, $NO_3^-$, $NH_4^+$ or $HCO_3^-$. We performed preliminary tests to add the exact concentrations of each of these products in tap water in order to find the chemical species actually involved.

Unfortunately, each product has a beneficial effect on growth as we had assumed ($H_2O_2$: +88%, $NO_2^-$: +62%, …) and even the temperature of the water after exposure can have an influence (+72%). Therefore, in order to be able to understand and discuss the actual mechanisms, the plan of experiment taking into account all species combinations are too important to be developed in this article.

Finally, as supported in Fig. 9b, the distributions of stem and root lengths is very spread out if one compares UTW and PATW. This observation seems to support that within a same seed family, the plasma activation does not act in the same way on each seed. This activation can be as effective as it is without having a drastic effect. It is therefore important to note that plasma activation can result in different responses both on the same batch of seeds as on seeds of different families. These response discrepancies could depend on seeds internal mechanisms, e.g. their sensitivity and selectivity thresholds to chemical species or to different defense mechanisms.

# 5. Conclusion

In this article, we have characterized and quantified 16 long lifetime chemical species in PATW. We have shown that the chemical composition of PATW has no impact on pH but a major one on electrical conductivity. Reactions between gas phase species and tap water lead to the formation of aqueous species like hydrogen peroxide, nitrite, nitrate or ammonia. Conversely this interaction activates the consumption of bicarbonates ions.

Due to acid/base equilibria and the increase of water temperature during plasma activation, a redistribution of chemical species is observed from their basic to acid form. We have also demonstrated that variations of electrical conductivity during activation are thoroughly described by the variation of two predominant ions: nitrate and bicarbonates ions.

In addition, we have shown that placing the plasma device in a confined enclosure modified the chemical reactions in gas phase and therefore in the liquid. Initially filled with air, the plasma activation increases relative humidity and ozone concentrations in the enclosure. The increase of relative humidity favors the production of hydroxyl radical and consequently of aqueous hydrogen peroxide. These reactions result in a quadratic and non-linear increase in the concentration of hydrogen peroxide in PATW as a function of the activation time. A higher concentration of ozone combined with a decrease of molecular oxygen in the reactor enclosure lead to a lower production of nitric oxide over time and finally of nitrites and nitrates in the PATW.

Concerning the agronomical part of this work, a proof of concept has been demonstrated in which plasma-activated tap water for 15 min has shown significant and positive effects on the growth of coral lentils with increase in plant length after 6 days of PATW activation (+128%).

From the state of art on chemical treatment of seed, we have been able to demonstrate the possible contribution of five chemical species on the growth of plants by DBD activations. Formation of aqueous nitrite, nitrate, ammonium ions and hydrogen peroxide during treatment reaches concentrations such that each one can induce an increase in growth. Conversely, the consumption of bicarbonate ions during plasma activation can also promote seedlings growth.

# Acknowledgements

This work has been achieved within the LABEX Plas@Par project, and received financial state aid managed by the Agence Nationale de la Recherche, as part of the Programme Investissements d'Avenir (PIA) under the reference ANR-11-IDEX-0004-02. Also, this work was supported by grant number 265356 from CNRS (Mission pour l'interdisciplinarité). The authors thank P. Auvray for assembling the AC generator.